\documentclass[journal]{IEEEtran}

\IEEEoverridecommandlockouts
\usepackage{cite}
\usepackage{booktabs}
\usepackage{amsmath,amssymb,amsfonts}
\usepackage{algorithmic}
\usepackage{graphicx}
\usepackage{subfigure}
\usepackage{textcomp}
\usepackage{tikz}
\usepackage{xcolor}
\usepackage{pgfplots}
\usepackage{pgfplotstable}
\usepackage{tikz,pgfplots}
\usetikzlibrary{positioning,shadows,shapes,backgrounds,calc}
\usepackage{pgfplotstable}
\pgfplotsset{compat=1.12} 

\usepackage{ragged2e}
\usepackage{booktabs,makecell,multirow,tabularx}
\usepackage{wrapfig}
\usepackage{dblfloatfix}
\usepackage{blindtext}
\usepackage{cuted}
\usepackage{subfigure}
\usetikzlibrary{spy}
\usepackage[pdftex]{hyperref} 
\hypersetup{colorlinks,linkcolor=red,citecolor=blue} 
\usetikzlibrary{shapes.geometric}
\usetikzlibrary{shapes}
\usetikzlibrary{fit}
\usetikzlibrary{shapes.multipart}
\usetikzlibrary{positioning}
\pgfplotsset{compat=1.17}
\usetikzlibrary{fadings}

\usepackage{balance}
\usepackage{pifont}
\usepackage{bbding}
\usepackage{fontawesome}

\newcounter{lemma}

\newtheorem{exampleplain}{Example}
\newtheorem{remark}{Remark}
\newenvironment{example}{\begin{exampleplain}}{~\hfill$\vartriangle$\end{exampleplain}}

\definecolor{orange}{rgb}{1,0.7,0}
\definecolor{myDarkGreen}{rgb}{0.00000,0.58824,0.00000}%

\DeclareMathOperator*{\argmax}{argmax}

\newcommand{\bB}{\boldsymbol{B}}

\newcommand{\bX}{\boldsymbol{X}}\newcommand{\bx}{\boldsymbol{x}}
\newcommand{\bY}{\boldsymbol{Y}}\newcommand{\by}{\boldsymbol{y}}

\newcommand{\bb}{\boldsymbol{b}}
\definecolor{myred}{RGB}{255,0,0}

\newcommand{\changed}[1]{\textcolor{black}{#1}}
\newcommand{\ch}[1]{\textcolor{black}{#1}}
\definecolor{myGreen}{rgb}{0.00000,0.49804,0.00000}

\definecolor{yellow}{RGB}{250, 199, 100} 
\definecolor{blue}{rgb}{0.38, 0.51, 0.71} 
\definecolor{darkblue}{RGB}{17, 42, 60} 
\definecolor{red}{RGB}{175, 49, 39} 

\definecolor{orange}{RGB}{217, 156, 55} 
\definecolor{green}{RGB}{144, 169, 84} 
\definecolor{palegreen}{RGB}{197, 184, 104} 

\definecolor{yellow}{RGB}{250, 199, 100} 
\definecolor{brokenwhite}{RGB}{218, 192, 166} 
\definecolor{brokengrey}{rgb}{0.77, 0.76, 0.82} 

\usetikzlibrary{patterns}

\begin{document}
\title{
On Shaping Gain of Multidimensional Constellation  in Linear and Nonlinear Optical Fiber Channel}

\author{Bin~Chen,~\IEEEmembership{Senior Member,~IEEE}, Zhiwei~Liang,~\IEEEmembership{Student Member,~IEEE}, Yi~Lei,~\IEEEmembership{Member,~IEEE}, 
JingXin~Deng,~\IEEEmembership{Student Member,~IEEE},
Shen~Li,~\IEEEmembership{Member,~IEEE}, and Gabriele~Liga,~\IEEEmembership{Member,~IEEE}

\thanks{The work of B.~Chen, Y.~Lei, Z.~Liang and J.~Deng are supported by the National Key Research and Development Program of China (No.~2024YFB2908400),  the National Natural Science Foundation of China  (No.~62171175 and 62001151) and the Fundamental Research Funds for the Central Universities under Grant JZ2024HGTG0312. 
The work of S.~Li was  supported by the Swedish Research Council (No. 2021-03709).
Parts of this paper have been presented at the \textit{European Conference Optical Communication (ECOC)}, Glasgow, Scotland, September 2023 \cite{BinChenECOC2023}.}

\thanks{B.~Chen, Y.~Lei, Z.~Liang, and J.~Deng  are with  School of Computer Science and Information Engineering, Hefei University of Technology, Hefei, China and also with Intelligent Interconnected Systems Laboratory of Anhui Province (e-mails:~\{bin.chen,leiyi\}@hfut.edu.cn).
}
\thanks{S.~Li was with Department of Electrical Engineering, Chalmers University of Technology, Gothenburg, Sweden. She is  now  with Centre of Optics, Photonics and Lasers (COPL), Department of Electrical and Computer Engineering, Université Laval, Québec, Canada (e-mail:~shen.li.1@ulaval.ca).
}

\thanks{G.~Liga is 
with  Department of Electrical Engineering, Eindhoven University of Technology, Eindhoven, The Netherlands (e-mail:~g.liga@tue.nl).
}

}

\maketitle
\begin{abstract}
Utilizing the multi-dimensional (MD) space for constellation shaping has been proven to be an effective approach for achieving shaping gains.  Despite there exists a variety of MD modulation formats tailored for specific optical transmission scenarios, there remains a notable absence of a dependable comparison method for efficiently and promptly re-evaluating their performance in arbitrary transmission systems. In this paper, we introduce an analytical nonlinear interference (NLI) power model-based shaping gain estimation method to enable a fast performance evaluation of various MD modulation formats in coherent dual-polarization (DP) optical transmission system. In order to extend the applicability of this method to a broader set of modulation formats, we extend the established NLI model to take the 4D joint distribution into account 
and thus able to analyze the complex interactions of non-iid signaling in DP systems. 
With the help of the NLI model, we conduct a comprehensive analysis of the state-of-the-art modulation  formats 
and investigate their actual shaping gains in two types of optical fiber communication scenarios (multi-span and single-span).    
The numerical simulation shows that for arbitrary modulation formats, the NLI power and relative shaping gains in terms of signal-to-noise ratio can be more accurately estimated by capturing the statistics of MD symbols. 
Furthermore, the proposed method further validates the effectiveness of the reported NLI-tolerant modulation format in the literature, which reveals that the linear shaping gains and modulation-dependent NLI should be jointly considered for nonlinearity mitigation.
\end{abstract}


\section{Introduction}
Through a series of revolutionary technological advances,  optical transmission systems have enabled the growth of Internet traffic for decades \cite{David2010}.  The majority of the extensive bandwidth available in fiber systems is already being utilized, and as a result, the capacity of the optical core network cannot keep pace with the continuous growth of traffic \cite{Essiambre2012}. In order to meet the growing demand, it is crucial to increase the spectral efficiency (SE) in optical fiber communications. In wavelength division multiplexing (WDM) transport systems with coherent detection, high-order modulation formats, such as polarization multiplexed $M$-ary quadrature amplitude modulation formats (PM-$M$QAM), have been widely used to improve the SE \cite{EssiambreJLT2010}. 
However, as the modulation order increases, the maximum rate that can be achieved with uniform signaling starts to suffer from a loss with respect to the capacity of the additive white Gaussian noise (AWGN) channel, i.e., the shaping gap \cite{ForneyJSAC1984}.
Therefore, numerous constellation shaping techniques that construct a signal with a Gaussian-like distribution have been widely investigated in an attempt to close the shaping gap. 
 
Constellation shaping has been also extensively studied in fiber optic communications to increase the achievable information rates (AIRs), including probabilistic shaping (PS) \cite{Kschischang1993, Ghazisaeidi2016}, geometric shaping (GS) \cite{ForneyJSAC1984, Galdino2020} and the combination of these two techniques \cite{8346117,Cai17OFC,Oliari2022,CaiJLT2020}. 
In the former, the constellation points presents nonuniform probability distribution. This can provide near-optimal linear shaping gain through compatibility with traditional QAM formats.  In geometric shaping, the probability of the constellation points remains uniform while altering their position. \changed{Note that the projection of   multi-dimensional geometric shaping  on lower dimension (e.g. in 2D) is also nonuniform probabilistic distributed.} 
For the AWGN channel, both PS and GS can achieve the Gaussian capacity distribution to close the shaping gap. 
Via GS and PS an optimal shaping distribution tailored for varying signal-to-noise ratios (SNRs) can also be attained. 
In addition, combining PS and GS as a hybrid shaping was also demonstrated to outperform using GS alone in optical fiber channel \cite{Oliari2022}. 
However, designing and implementing PS or/and GS brings additional challenges to the communication system. 
PS requires an efficient implementation of the amplitude (de)shapers with finite block length while the rate losses incurred by distribution matching (DM) encoders significantly diminish the benefits of PS \cite{Fehenberger2020JLT}. The GS requires the design and implementation of low-complexity (multidimensional) mappers/demappers \cite{Yoshida2016ECOC}. 
Additionally, both PS and GS cannot guarantee to achieve optimal performance for practical channels such as fiber optic nonlinear channels, necessitating the optimization of constellations tailored to these specific nonlinear channels \cite{Liu2023,Ling2022}. 
To maximize the benefits from shaping, modulation formats with higher number of dimensions  can be increased to have a greater impact on system performance. Thus, the design of multi-dimensional (MD) modulation formats has been considered as an effective approach to harvest performance gains in optical communications.

The application of MD modulation in optical fiber communications has already undergone extensive research \cite{Chagnon:13,Kojima2017JLT,BinChenJLT2019,SebastiaanJLT2023,GabrieleOFC2022,EricJLT2022,Shiner:14,El-RahmanJLT2018,Mirani2021,Li2023}. In coherent fiber communication, there are essentially four dimensions, namely two orthogonal quadratures in two orthogonal polarizations.  By considering joint modulation of these four dimensions, greater performance can be achieved. The utilization of this concept in optical communications may be traced back to the first generation of coherent optical communications research in the 1980s \cite{Betti1990}, who exploits these four degrees of freedom of the electromagnetic wave propagating through the fiber to get closer to the theoretical Shannon limit.  The concept was more recently re-introduced by performing constellation shaping jointly on these four dimensions, which can achieve higher shaping gains or coding gains attracting particular interest from the research community. 
In addition, research has also been conducted into the design of modulation formats to expand the number of modulation dimensions beyond four, utilizing additional dimensions such as time slots, wavelengths, and spatial dimensions \cite{Shiner:14,El-RahmanJLT2018,Mirani2021,BinChenPTL2019,Li2023}. For example, the eight-dimensional formats were obtained in two consecutive time slots in \cite{Bendimerad:18}.  A 12D modulation format was demonstrated across three linearly coupled spatial modes of a multicore fiber \cite{ReneOFC2020}. In \cite{Dar14_ISIT,OmriJLT2016}, MD modulation formats have been shown to reduce the NLI generated by reducing the variations of the transmitted signal energy in addition to the usual (linear) shaping gain, and it could potentially be higher than the ultimate shaping gain (1.53~dB) in AWGN channel.

Since shaping in MD space can effectively mitigate nonlinear effects, more researchers have focused on the search for modulation formats in higher dimensions that can tolerate linear noise and/or nonlinear interference \cite{Dar14_ISIT} and various heuristic ideas have been proposed to improve nonlinear tolerance and reduce optimization complexity \cite{Kojima2017JLT,BinChenJLT2019,SebastiaanJLT2023,Li2023}. For example, the 4D-PRS64 \cite{BinChenJLT2019} or 4D2A8PSK \cite{Kojima2017JLT} format was obtained under the constraint of the 4D constant modulus, which showed great nonlinearity tolerance performance because of the less energy variations. The 4D-OS128 format \cite{BinChenJLT2021} was obtained by maximizing generalized mutual information (GMI) with orthant-symmetry idea which can significantly reduce the dimensionality of searching space. In \cite{SebastiaanJLT2023}, shell shaping was introduced as an approach to close the nonlinearity-caused shaping gap. Furthermore, a low-complexity Voronoi constellations with a cubic coding lattice of up to 32 dimensions were proposed in \cite{Li2023}. 
Despite there are various of existing MD modulation formats in the literature, which are designed for AWGN channel or a specific optical transmission scenario, it is still lack of a reliable comparison method for effectively and quickly re-evaluating their performance in various optical transmission systems.

To address the above problem, we introduce a new approach to quickly evaluate the actual shaping gains of MD modulation formats in different transmission systems. The main contributions in this manuscript can be summarized as below:
\begin{itemize}
    \item We present a fast  shaping gain estimation approach for MD modulation formats in optical fiber channel supported by the use of an NLI analytical model.
    The NLI model can provide easy-to-compute NLI power as well as accurate effective SNR for all the possible MD dual polarization (DP) constellations, including advanced joint 4D constellation and independent 4D constellation over different modes, such as multi-mode and multi-core transmission systems. 
    \item 
    To make this method applicable to MD modulation formats, we extend our previous work in \cite{2020Extending,Liang2023JLT} to analyze the complex interactions of non-i.i.d. signaling in dual-polarization systems. 
    According to the expression of NLI power, the NLI features are further analyzed.
    \item According to the proposed shaping gain estimation approach, we comprehensively analyzed the performance of various MD modulation formats in both the AWGN channel and the optical fiber channel. 
    The results show that the overall performance of MD formats should be evaluated by combing the shaping gains in the linear channel and the modulation-dependent effective SNR loss in the optical fiber channels. 
    The results in this work confirm once again that the constellations with less energy variations can trade-off the linear shaping gains and nonlineariry tolerance for the specific optical transmission scenario. 
\end{itemize}

\begin{figure*}[!tb]
    \centering
    \includegraphics[height=20em]{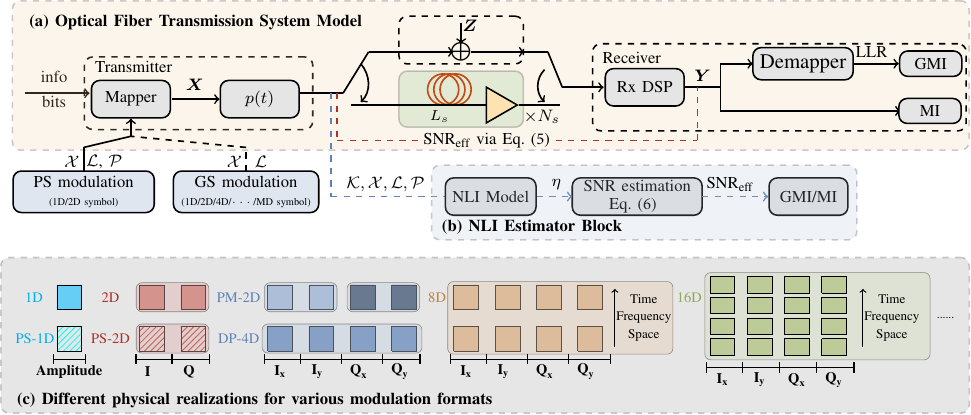}
    \caption{System model under consideration in this manuscript: (a) the block diagram of the  considered generic  optical fiber system employing various modulation format, (b) the nonlinear interference  estimation block for a given modulation format with its shape and distribution $(\mathcal{X}, \mathcal{L}, \mathcal{P})$ over a given optical fiber transmission link with parameters $\mathcal{K}$, and \changed{(c) an illustration to indicate the employed modulation formats, including 1D, 2D, 4D, 8D, and 16D with either uniform (filled squares) or non-uniform (squares with slashes) probability distribution}. Each colored square denote a one-dimensional real signal space, which is most intuitively viewed as a real-valued amplitude symbol. {1D and 2D Formats with independent symbols are shown in as separate one-square block and two-squares block, respectively.}}
    \label{fig:System_Model}
\end{figure*}

The manuscript is organized as follows: in Section~\ref{sec:system},  we  present the system model and review the performance metrics, including mutual information (MI), GMI and effective SNR for MD modulation formats. Section~\ref{sec:NLI_model} presents the key expression of the proposed NLI model and validation of the 4D NLI model.  In Section~\ref{sec:results}, we analyze the performance of various MD modulation formats in both the AWGN channel and the optical fiber channel. In Section~\ref{sec:discussion}, we discuss the physical realizations and digital signal processing of various MD formats. Finally,  \changed{Section~\ref{sec:conclusion} concludes this manuscript.}

\section{System Model and  Performance Metrics}\label{sec:system}
In this section, we introduce the system model that we consider for optical transmission and metrics for performance evaluation. We also briefly discuss the various options for mapping MD symbols to signal components in a dual-polarization optical fiber transmission system.

\subsection{System Model}\label{sec:system_model}
Fig.~\ref{fig:System_Model} shows the system model for optical fiber system and the corresponding schematic diagram of the fast evaluation method. Fig.~\ref{fig:System_Model} (a) represents a generic multi-span \changed{WDM}  optical fiber communication system. 
To simulate the performance of various constellations with different constellation cardinalities in a linear or nonlinear transmission scenario, the cases of $N_s$-span (optical fiber channel) and AWGN channel are considered. At the transmitter side, we predefine the MD modulation formats (including regular QAM and advanced MD formats) or PS-QAM to be transmitted, where $\mathcal{X}$, $\mathcal{L}$ and $\mathcal{P}$ are constellation coordinates, the corresponding binary labeling and probability distribution, respectively. \ch{The input information bits are mapped into MD symbols $\boldsymbol{X}$ with $N$ real dimensions drawn  from a discrete  MD modulation format $\mathcal{X}$ with cardinality $M=2^m=|\mathcal{X}|$ and its corresponding binary labeling $\mathcal{L}$ via a mapper.} 
Furthermore, the transmitted symbols are pulse-shaped by a real pulse $p(t)$. Then the resulting MD symbols are transmitted through the AWGN channel or optical fiber channel. In the AWGN channel scenario, the additive Gaussian noise $\boldsymbol{Z}$ is added.  The optical fiber channel is an $N_s$ fiber link where each span has a span length of $L_s$ and each is followed by an ideal lumped optical amplifier for which the gain exactly recovers the span losses.  At the receiver, the received symbols are processed by a receiver digital signal processing (DSP) block, including ideal chromatic dispersion compensation (CDC), matched filtering, sampling and ideal phase compensation for potential constant phase rotation. The output symbols $\boldsymbol{Y}$ are then processed by a demapper to generate soft information (i.e., log-likelihood ratios), which is used to finally estimate the transmitted information bits and compute the performance metrics of the system.

As shown in Fig.~\ref{fig:System_Model} (b), the nonlinear interference estimation block is introduced as a low-complexity alternative. By inputting predefined MD modulation format $\mathcal{X}$ and a probability distribution $\mathcal{P}$ and the given channel parameters $\mathcal{K}$ (e.g. fiber parameter and link parameter) to an analytical NLI model, we can obtain the nonlinear power coefficients $\eta$ which can be used to calculate the effective SNR through Eq.~\eqref{SNR-general2} \changed{(defined in Sec.~\ref{sec:SNR})}. Subsequently, the constellation $\mathcal{X}$, \changed{the binary label $\mathcal{L}$} and the effective SNR associated with a given channel law are used as input parameters to estimate the achieve information rate (AIR: MI or GMI) \cite{AlvaradoJLT2018}.

In this work, we study \changed{two classes of} modulation formats: MD modulation formats (GS in 2D, 4D, ..., 32D space) and PS-QAM formats.  It is well known that, in coherent fiber optical communications systems, the  information bits can be modulated through four degrees of freedoms (DOFs) in electromagnetic fields, namely amplitude (I quadrature), phase (Q quadrature), and x/y polarization. 
As shown in the Fig.~\ref{fig:System_Model} (c), we show the MD modulation with different dimension realizations. 
\changed{To provide a visual representation of the uniform MD formats and  PS formats, the filled square denotes that  all points in the constellation are equally probable. In contrast, the square with slashes denotes the constellation with unequal probability, which is either  independent (PS-1D) or joint modulated (PS-2D) over I and Q quadratures.} 
For 2D space, the transmitted symbols are jointly modulated over I and Q quadratures. 
\changed{For 4D space, two identical or different 2D formats are used to transmit information independently over the two orthogonal polarizations. This is the case of polarization-multiplexed 2D (PM-2D) constellations. To emphasize the possibilities of PM-2D constellations with two different 2D formats over the two orthogonal polarizations, two different  blue colors are used. For the dual-polarization 4D (DP-4D) constellations, the transmitted symbols  are jointly modulated in the 4D space, and thus are represented by using the same blue color.} 
Note that  the 4D NLI model \cite{2020Extending,Liang2023JLT} is equivalent to the EGN model for PM-2D constellations.  
For the MD modulation formats ($N >$ 4), the constellations are jointly shaped over multiple dimensions (e.g., time slots, frequency and space).  Further exploration of the practical implementation challenges arising from increasing dimensionality through additional physical DOFs are addressed in Section~\ref{sec:discussion}.

\subsection{Achievable Information Rates}\label{sec:AIR}
As the most popular information-theoretical performance metrics, mutual information, and generalized mutual information quantify the maximum number of information bits per transmit symbol in the symbols-wise and bit-wise coded modulation systems, respectively \cite{AlvaradoJLT2018}. 
In general, MI applies to nonbinary codes or multilevel codes with multi-stage decoder, while GMI is suitable for the bit interleaved coded modulation (BICM) system \cite{AlvaradoJLT2018,Chen2023OE}.  
\ch{Since both the uniform and non-uniform distributed constellations are considered in this work, the general AIRs calculation principles are reviewed.} 
\ch{For an arbitrary MD memoryless channel, the MI is given as \cite[Eqs.~(3-5)]{AlvaradoJLT2018}
\begin{align}\label{MI}   
\begin{split}
     \text{MI} &\triangleq \mathbb{E}\left[\text{log}_2\frac{P_{\bY|\bX}(\bY|\bX)}{ P_{\bY}(\bY)}\right]\\
     &=\frac{1}{M}\sum_{i=1}^M\int_{\mathbb{R}^N}P_{\bY|\bX}(\by_i|\bx_i)g_i(\by) d\by,
\end{split}
\end{align}
with
\begin{align}
    g_i(\by) = \text{log}_2\frac{P_{\bY|\bX}(\by_i|\bx_i)}{\sum_{j=1}^M P_{\bX}(\bx_j)P_{\bY|\bX}(\by_i|\bx_j)},
\end{align}
in bits/symbol per $N$ real dimensions, where $\mathbb{E}[\cdot]$ denotes the expectation,  the integral is an  integral over the $N$-dimensional real space, $M$ is the constellation cardinality size, $P_{\bX}(\bx_j)$ indicates the probability of transmitting $\bx_j$ and the $P_{\bY|\bX}$ is the ``matched" channel law given by \cite[Eq.~(19)]{AlvaradoJLT2018}.}

\ch{The MI is achievable for a demapper-decoder that uses the memoryless $P_{\bY|\bX}$. For the BICM scheme, the bit-wise demapper is followed by a binary decoder. In this setup, the symbol-wise input $\bX$ is fully determined by its binary representation $\bB = [B_1,..., B_m]$, where each bit level $\bb_l = [b_{1,l},b_{2,l},...,b_{m,l}]$ can be stochastically dependent. Therefore, the bit-wise decoding rule is not same as the symbol-wise decoding rule. Generally, the bit-wise decoder can be cast into the framework of a mismatched decoder, and thus the GMI for MD memoryless channel is given as \cite[Eqs.~(12,13)]{8640810}}
\ch{\begin{align}\label{GMI}
\begin{split}
    \text{GMI} & \triangleq \max_{s\geq 0} \mathbb{E}\left[\text{log}_2\frac{q(\bB, \bY)^s}{\sum_{\bb\in \{0,1\}^m} P_{\bB}(\bb)q(\bb,\bY)^s }\right]\\
    &= \max \{0, H(\bB) - \sum_{i=1}^m H(B_i|\bY)\},     
\end{split}
\end{align}
in bits/symbol per $N$ real dimensions, where $H(\cdot)$ denotes entropy and $s$ is a nonnegative scaling parameter. The $q(\bB, \bY)$ is the mismatched decoding metric and $P_{\bB}(\bb)$ is the bit distribution.}  

\ch{In order to make a fair comparison for arbitrary $N$-dimensional modulation formats, we can derive an appropriate normalization of AIRs (NMI or NGMI), which serve as a channel metric quantifying the number of information bits per transmitted bit. For arbitrary MD modulation formats, the NMI or NGMI can be given as \cite[Eq.~(14)]{8640810}
\begin{align}
    R^* = 1 - \frac{H(\bX) - R}{m},
\end{align}
where $R$ is equal to MI or GMI, and the units of $H(\bX)$ is bits/symbol per $N$ real dimensions.  Note that the normalized AIRs for probabilistically-shaped signaling were defined as asymmetric information (ASI) in \cite[Eq.~(11)]{8076914}, which has the same form as $R^*$. In addition, for the finite-length distribution matcher, the DM rate loss $R_L$ should be considered into the conventionally estimated AIR \cite[Eq.~(8)]{8850066}, i.e., $\text{AIR}_{\text{DM}} = \text{AIR} - R_L$.}  
The values of NMI and NGMI indicate the largest code rate that is able to guarantee error-free post-FEC results for the symbol-wise and bit-wise coded modulation system, respectively. 

\subsection{Effective Signal-to-Noise Ratio}\label{sec:SNR}
As shown in the Fig.~\ref{fig:System_Model} (a) and (b), the effective SNR can be estimated using a numerical computation method or utilizing the NLI power coefficient $\eta$ obtained from the analytical NLI model. For the optical fiber transmission scenario, the impairments arising from the optical amplifiers and fiber nonlinearity are taken into account. Assuming that all of these impairment factors can be modeled as statistically independent additive noise sources, the effective SNR (the received signal after fiber propagation and the receiver digital signal processing including chromatic dispersion compensation and phase compensation) can be expressed as \cite{Poggiolini_JLT2014, BinChenJLT2019}
\begin{align}
    \text{SNR}_{\text{eff}} &=\frac{\mathbb{E}[||\boldsymbol{X}||^2]}{\mathbb{E}[||\boldsymbol{Y}-\boldsymbol{X}||^2]} \label{SNR-general1}\\
    &=\frac{P}{\sigma_{\text{ASE}}^2 + \sigma_{\text{NLI}}^2 }= \frac{P}{\sigma_{\text{ASE}}^2 + \eta P^3}\label{SNR-general2},
\end{align}             
where $P$ is the optical launch power, the $\sigma_{\text{ASE}}^2$ denotes the  amplified spontaneous emission (ASE) noise power which generates from the optical amplification process. The $\sigma_{\text{NLI}}^{2}$ denotes the NLI noise power, including both intra- and inter-channel distortions. The $\eta$ is the functions of several different intra- and cross-polarization moments of a transmitted modulation format. In Eq.~\eqref{SNR-general1}, the effective SNR is estimated via the Split-step Fourier Method (SSFM), which is computationally expensive and time-consuming. In Eq.~\eqref{SNR-general2}, the effective SNR can be rapidly evaluated by the NLI model, and thus avoiding a large number of Fourier transform processes. \changed{Note that in practical systems, the transceiver noise and polarization-dependent distortions can be dominant  and influence the entire analysis. Especially, polarization-dependent distortions, including polarization mode dispersion (PMD), polarization-dependent loss (PDL), and states of polarization (SOP), can break the energy-constant feature of the multidimensional constellations. Meanwhile, these distortions also provide a benefit for MD modulation design to mitigate the interference between different polarizations.}

MD constellation shaping aims at reducing the required SNR over the AWGN channel but in practice often leads to a smaller received (effective) SNR due to hardware distortions and fiber nonlinearities. Thus, a true performance improvement of MD formats in the optical channel is achieved if and only if the reduction in the required SNR is larger than the loss in terms of the effective SNR. In the remainder of this manuscript, the required SNR, which is defined as the minimum SNR to achieve a target value of NMI/NGMI, is used as a performance metric to evaluate the tolerance to nonlinearity. 

\section{NLI model for Arbitrary Dual-Polarization 4D Modulation Formats}\label{sec:NLI_model}
In this section, we derived an analytical expression that enables the rapid calculation of the nonlinear interference power. According to our recently proposed NLI model, we derive the approximation of the modulation-independent and modulation-dependent  nonlinear interference power as a function of fiber-link parameters. Then we discuss using the model to evaluate the shaping gain for the nonlinear optical channel and design the nonlinearity-tolerant constellation.

\subsection{The Main Formulas of the 4D NLI  Model}
 In the classic GN model of \cite{Dar:13}, the derivation of the NLI power coefficient is based on the assumption that the signal statistically behaves as Gaussian noise over uncompensated links. Therefore, the choice of transmission modulation formats in the GN model has no effect on the NLI noise power. To analyze and reduce the limitations of the GN model, the EGN model has been presented in \cite{Carena:14}, which lifts the Gaussianity assumption and proposed several modulation-dependent correction terms. For PM-2D modulation formats, the EGN model has been demonstrated to possess high accuracy. However, the EGN model is only valid for PM-2D modulation formats in which polarizations act as two independent channels.

In order to make the NLI model suitable for the DP-4D formats, the analytical 4D NLI model proposed in \cite{Liang2023JLT} without any assumptions on either the marginal or joint statistics of the two polarization components of the transmitted 4D modulation formats besides being zero mean. In other words,  the simplified assumption that transmitted symbols are \changed{independent and identically distributed (i.i.d.) for x and y polarizations}, which was made in the EGN model, was removed.

\begin{figure}
    \centering
    \hspace{-1em}
    \includegraphics[width=.5\textwidth,height=17em]{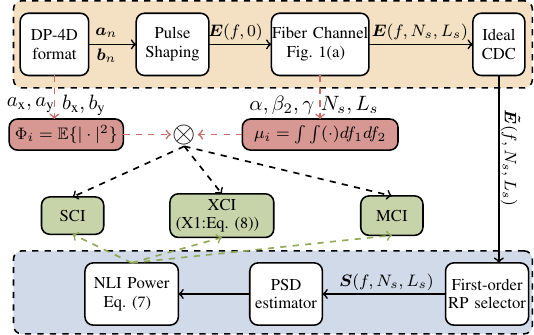}
    \caption{The main derivation steps of the 4D NLI model which consists of three main NLI. The orange block represents the  optical fiber system model under investigation and the blue block represents the detail derivation of the  4D NLI power  estimation module in Fig.~\ref{fig:System_Model} (b).
    }
    \label{fig:NLI_Model}
        \vspace{-0.5em}
\end{figure}

The main analytical derivation of the 4D NLI model is shown in  Fig.~\ref{fig:NLI_Model}.
In this manuscript, it is assumed that a multi-channel signal is transmitted. Thus, the same modulation format is modulated to different WDM channels. 
After pulse shaping and multiplexing, the $\boldsymbol{E}(f,0)$ is the Fourier transform of transmitted signal $\boldsymbol{E}(t,0)$, which is transmitted over optical fiber channel with $N_s$ fiber spans. Note that here we only focus on the signal-signal nonlinear interference, and thus, the signal-noise nonlinear interference added by the optical amplifier was entirely ignored. 
As is well known, the Manakov equation plays a significant role in understanding fiber propagation effects and is crucial for deriving an analytical expression of NLI power.  Here, we solve the Manakov equation within the framework of first-order perturbation theory, expressed as $\boldsymbol{E}(f,N_s,L_s)$.  Generally, we are more concerned with the field after ideal CDC, and thus, the exponential $e^{j2\beta_2\pi^2f^2N_sL_s}$ can be removed from $\boldsymbol{E}(f,N_s,L_s)$. The signal after CDC is represented by $\boldsymbol{\tilde{E}}(f,N_s,L_s)$. Then we isolate the first-order  regular perturbation (RP) term $\boldsymbol{\tilde{E}}_1(f,N_s,L_s)$ in the first-order RP selector block and compute its power spectral density (PSD) $\boldsymbol{S}(f, N_s,L_s)$ in the PSD estimator block. Finally, the NLI power $\sigma^2_{\text{NLI}}$ for both x and y polarizations in Eq.~\eqref{SNR-general2} can be expressed as
\begin{align}\label{eta}
\begin{split}
    \sigma^2_{\text{NLI}} = [\sigma^2_{\text{NLI,x}}, \sigma^2_{\text{NLI,y}}]^T =& \Bigg[\int_{-\infty}^{\infty}S_{\text{x}}(f,N_s,L_s)|P(f)|^2df,\\
    & \int_{-\infty}^{\infty}S_{\text{y}}(f,N_s,L_s)|P(f)|^2df\Bigg]^T,
\end{split}
\end{align}
where the $P(f)$ is the transmitted pulse spectrum which assumed to be strictly band-limited. 
The NLI power in this manuscript contains three parts, including self-channel interference (SCI), cross-channel interference (XCI) and  multiple-channel interference (MCI) \cite[Eq.~(13),(16),(18)]{Liang2023JLT}. In order to further analyze the characteristics of the NLI noise and understand the impact of modulation formats on Kerr nonlinearity, we take the cross-phase modulation (XPM) as an example.\footnote{There are four terms in XCI, called as X1, X2, X3 and X4, where the X1 corresponds to XPM as shown in \cite[Fig.~2]{Liang2023JLT}.} The XPM nonlinear power $\sigma_{\text{X1,x}}^2$ for the x polarization can be written as
\begin{align}\label{X1}
\begin{split}
    \sigma_{\text{X1,x}}^2 & = \underbrace{(4\mu_1 - 8 \mu_3)\Phi}_{\text{Modulation-independent}} + \underbrace{[\Phi_1\mu_1 + \Phi_2\mu_2 + \Phi_3\mu_3]\Phi}_{\text{Modulation-dependent}}.
\end{split}
\end{align}  
where the $\Phi = \mathbb{E}\{|a_\text{x}|^2\}\mathbb{E}^2\{|b_\text{x}|^2\}$, in which the \changed{$a_\text{x}$ and $b_\text{x}$} are random variables in x polarization, transmitted by channel of interest and interference channel, respectively. 
The $\Phi_1, \Phi_2, \Phi_3$ are functions of several different modulation-dependent intra- and cross- polarization moments  and the $\mu_1, \mu_2, \mu_3$ are the integrals related to channel parameters. For more details about these coefficients, we refer the reader to  \cite{2020Extending, Liang2023JLT}.

According to Eq.~\eqref{X1}, two important features of NLI can be identified. The first feature is its dependence on the average power of the transmitted signals.  Specifically, Eq.~\eqref{X1} indicates that the NLI power is proportional to $\Phi$, where $\Phi$ contains the average symbol energy of the channel of interest (COI) in x polarization ($\mathbb{E}\{|a_\text{x}|^2\}$) and the average symbol energy of the interfering (INT) channel in x polarization ($\mathbb{E}\{|b_\text{x}|^2\}$), respectively. Assuming that the transmission power of both COI and INT channel in x polarization  $P_\text{x}$ is same, the $\Phi$ is equal to $P_\text{x}^3$.\footnote{Note that the $P_\text{x}$ is no longer strictly equal to a half of total transmission power per channel $\frac{P}{2}$ for asymmetric DP-4D modulation formats.} Therefore, the NLI power is negligible at a low average signal power, but becomes the dominant effect at a high power. The second feature is the dependence on the modulation formats. As shown in the second part of Eq.~\eqref{X1}, the modulation-dependent noise power induced by the three terms $\Phi_1, \Phi_2, \Phi_3$. These terms are the forth- or sixth-order moment dependent on modulation formats.

All main nonlinear interference terms (SCI, XCI and MCI) were considered so that this model can provide reliable predictions of NLI for both low and high fiber dispersion regimes. On the other hand, its fast and accurate characteristics make it as a powerful tool for offering the potential for improved performance analysis in optical communication systems employing advanced modulation schemes.  
Therefore, this model represents a significant step forward in accurately representing the complex interactions of non-iid signaling in dual-polarization systems and helps to improve performance of achievable information rate by optimizing the MD constellation shaping, i.e., NLI model-aided constellation shaping. 
The application of this model will be introduced in subsection~\ref{sec:application_4D_NLI_model} and the accuracy will be further verified in subsection~\ref{sec:accuracy}.

\subsection{The Application of the 4D NLI Model}\label{sec:application_4D_NLI_model}
The application of this 4D NLI model is twofold. First, the 4D NLI model can be used as a low-complexity alternative to obtain NLI power of a given optical fiber channel. As shown in Eq.~\eqref{X1}, the analytical NLI model consists of integral terms and modulation-dependent coefficients, where all the integral terms can be calculated via either double or triple integrals, and the modulation-dependent coefficients $\Phi_1, \Phi_2, \Phi_3$ can be easily calculated for arbitrary 4D modulation formats. It suggests that the computational complexity of the 4D NLI model is mainly derived from the integral terms.  As demonstrated in \cite[Appendix~C]{Carena:14}, all triple or quadruple integrals can be simplified to require only a double integral for evaluation. Consequently, \changed{this 4D NLI model yields} achieved vast reductions in complexity compared to SSFM. This capability is crucial for efficient system design, optical networking that takes into account the physical layer, and real-time system optimization.  In addition, analytical models provide valuable insight into the dependencies between various parameters. For example, Eq.~\eqref{X1} shows the effect of the system parameters and the modulation format features on the NLI power.

Secondly, to design nonlinearity-tolerant modulation format in an efficient way, it is crucial to incorporate accurate knowledge of modulation-dependent nonlinear interference into the optimization process.  The impact of modulation formats is typically performed using the split-step Fourier method or NLI power models.  Therefore, the 4D NLI model is used to capture the impact of modulation formats and provide a cost function for modulation scheme optimization design. \changed{An MI} or GMI-based constellation optimization shaping of 4D formats in the nonlinear fiber channel can be defined as 
\begin{align}\label{opt_mi}
 \{\mathcal{X}^*\} &=\argmax_{\mathcal{X}:E[||X||^2]\leq \sigma^2_x} \{\text{MI}(\mathcal{X}, \mathcal{P}, \text{SNR}_{\text{eff}}^{\text{opt}})\}
\end{align}
and
\begin{align}\label{opt_gmi}
 \{\mathcal{X}^*,\mathcal{L}^*\} &=\argmax_{\mathcal{X},\mathcal{L}:E[||X||^2]\leq \sigma^2_x} \{\text{GMI}(\mathcal{X},\mathcal{L}, \mathcal{P}, \text{SNR}_{\text{eff}}^{\text{opt}})\},
\end{align}
where $\mathcal{X}^*$ and  $\mathcal{L}^*$ indicate the optimized constellation and labeling, respectively. The $\text{SNR}_{\text{eff}}^{\text{opt}}$ denotes the effective SNR in the optimum launch power $P^{\text{opt}}$, indicating the knowledge of modulation-dependent NLI. $P^{\text{opt}}$  is calculated by setting $\frac{\partial \text{SNR}_{\text{eff}}}{\partial P}$ to zero. Note that Eq.~\eqref{opt_mi} and Eq.~\eqref{opt_gmi} are denoted as geometrical shaping optimizations when $\mathcal{P}$ is set to equal probability. However, when $\mathcal{P}$ is non-equal probability, these equations are suitable for probabilistic shaping optimization or hybrid GS/PS optimization. This effect of modulation optimization on the change in $\mathcal{X}$ and $\text{SNR}_{\text{eff}}$ will be reflected in the gradient of the GMI cost function when using end-to-end learning.

\subsection{The Accuracy of the NLI 4D Model}\label{sec:accuracy}
In this subsection, \changed{the numerical validation} of this 4D NLI model (Eq.~\ref{eta} and Eq.~\ref{X1}) is performed via the effective SNR (Eq.~\eqref{SNR-general1} and Eq.~\eqref{SNR-general2}) in an optical fiber channel for PM-2D formats and DP-4D formats.

\begin{figure}[!tb]
    \centering
    \includegraphics[height=17.5em]{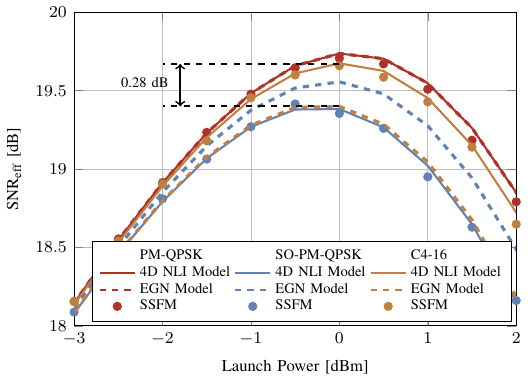}
    \vspace{-1em}
    \caption{The effective SNR as a function of launch power after 10 spans of SMF. The SSFM simulation results are marked as filled circles, while the results obtained from the NLI 4D  model, are marked as solid lines. The EGN model is denoted as dashed lines.}
    \label{fig:model_ssfm}
      \vspace{-0.5em}
\end{figure}

\begin{example}[The validation of the 4D NLI model]
In Fig.~\ref{fig:model_ssfm}, we show the effective SNR of three modulation formats through two analytical NLI models (4D NLI model and EGN model) and SSFM. The optical system under consideration is a 11-channel WDM  system with  45~GBaud symbol rate  and 50~GHz channel spacing. For the transmission link, a 10$\times$80~km multi-span standard single-mode fiber is used with loss coefficient $\alpha=$0.2~dB/km,  chromatic dispersion $D=$17~ps/nm/km, and nonlinear coefficient $\gamma=$1.3~/W/km. Each span is followed by an erbium-doped fiber amplifier with a noise figure of 5~dB. 
The modulation formats considered in our simulations are PM-QPSK and two 4D constellations, i.e., subset optimized PM-QPSK (SO-PM-QPSK) \cite{Martin2013LCOMM, Alex2015JLT} and C4-16 \cite{Alex2015JLT}. It can be observed that for \changed{PM-QPSK both EGN model} and 4D NLI model are in perfect agreement with the SSFM simulation results. For SO-PM-QPSK and C4-16, the 4D NLI model \changed{results  closely follow} the results obtained by SSFM.  In contrast, the EGN model fails to predict the effective SNR and  leads to the inaccuracies of  0.28~dB at the optimal launch power.  
\end{example}

The gap in Fig.~\ref{fig:model_ssfm} shows the obvious shortcoming of the EGN model in predicting the NLI of DP-4D modulation formats, and that the NLI can be underestimated (SO-PM-QPSK) or overestimated (C4-16). The main reason for this error is the inter-polarization dependency terms ignored in the EGN model. Note that for PM modulation formats such as PM-QPSK, the 4D NLI model yields the same results as the EGN model, since the new arising modulation-dependent corrections are equal to zero due to $\mathbb{E}\{a_\text{x}\} = \mathbb{E}\{a_\text{y}\}$. According to these results, the 4D NLI model demonstrates remarkable accuracy, which allows us to replace the time-consuming SSFM with this NLI model in subsequent simulations.

\section{Numerical Results}\label{sec:results}
In this section, we investigate the shaping gain of a series of multidimensional constellations in linear and nonlinear optical fiber channel. The multidimensional constellations which are under consideration in this manuscript are shown as follows:

\begin{itemize}
    \item Uniform PM-QAM formats.
    \item  MI-optimized or GMI-optimized 2D and 4D  formats.\footnote{The coordinates and bit-to-symbol mappings of the optimized constellation  are available at  \url{https://github.com/TUe-ICTLab/Binary-Labeling-for-2D-and-4D-constellations} \cite{Database-Github}.}
    \item MD Voronoi constellations (VCs) in \cite{Li2022, Li2023}, i.e., 8D-VC, 16D-VC and 32D-VC.
    \item Two nonlinearity-tolerant 4D formats (4D-64PRS \cite{BinChenJLT2019} and NLI-4D-1024 \cite{Ling2022}).\footnote{{Note that 4D-64PRS and NLI-4D-1024 is designed to be nonlinearity-tolerant by maximizing the GMI for a target SNR.}}
    \item \changed{PS-QAM ($n=10000$) is considered as the baseline with MI $=\{ 4.8, 6.4, 8\}$~bit/4D-sym, which probabilistically shapes PM-16QAM and PM-64QAM to an entropy of $H(\bX)= \{ 6.4, 8.8, 10.4\}$~bit/4D-sym based on CCDM with blocklength $n=10000$.}\footnote{\changed{When PS with blocklength $n=10000$, the correlation between symbols is negligible, thus it can be approximated via existing NLI models without considering a sliding window \cite{9464637}.}}
    \item \changed{PS-QAM ($n=100$) is realized by PM-16QAM and PM-64QAM constellations with finite shaping blocklength of $100$, 
    by employing the CCDM with the same entropy setup as PS-QAM ($n=10000$).}
\end{itemize}

\subsection{Numerical AWGN Simulations: Shaping Gap to AWGN Capacity}
In this subsection, we perform extensive MI and GMI evaluations for the considered formats in the AWGN channel. 
For the AWGN channel, the performance of an ideally shaped QAM system matches that of an ideal system using Gaussian modulation, which
represents the case of a Gaussian input distribution (equivalent to the case of dimensionality $N\rightarrow\infty$) \cite{Dar14_ISIT}.

To quantify the gains offered by different modulations, we define the ``SNR gap to AWGN capacity" ($\Delta\text{SNR}$). The value of $\Delta\text{SNR}$ in dB measures how far a given symbol-wise or  bit-wise coded modulation scheme is operating from the Gaussian modulation at a target NMI or NGMI, which are defined as
\begin{align}
\Delta\text{SNR}^R_{\text{req}}=\text{SNR}_{\text{req}}^R-\text{SNR}_{\text{req}}^{C},
\end{align}
where the $R$ is MI or GMI and $C$ is Shannon capacity. \ch{In this paper, we focus on an ideal SD-FEC with 25\% overhead in the BICM system, i.e., $R^*=0.8$. Thus, the required SNR can be obtained at $R=\frac{4}{N}(H(\bX)-m(1-R^*))$ or $C=\frac{4}{N}(m R^*)$ in bits/symbol per four dimensions.} 

\begin{figure}[!tb]
    \centering
    \includegraphics[width=.5\textwidth]{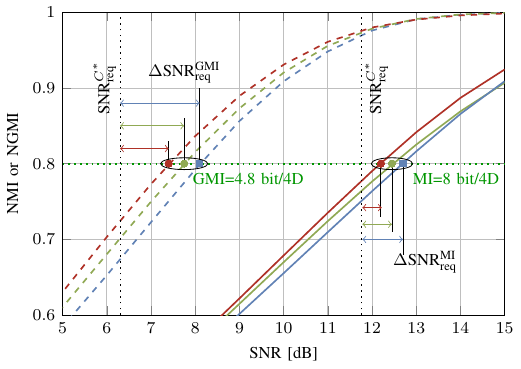}
       \vspace{-1em}
    \caption{Example of the ``SNR gap to AWGN capacity" for modulation formats with SE of 6~bit/4D and 10~bit/4D. The dashed green and red curves denote 2D and 4D GMI-optimal formats: DSQ2-8, GS-AWGN-4D-64, respectively. The solid green and red curves are 2D and 4D MI-optimal formats: GS-AWGN-2D-32 and C4-1024,  respectively. PM-QAM are shown as  baselines (blue curves). The black dotted line represent the  required SNR of the optimal Gaussian distributed modulation.}
    \label{fig:example}
     \vspace{-0.5em}
\end{figure}

\begin{example}[The SNR gap to AWGN capacity]
In Fig.~\ref{fig:example}, the linear performance in terms of NMI and NGMI for modulation formats with SE$={6, 10}$~bit/4D-sym is shown. The required SNR of Gaussian modulation $\text{SNR}_{\text{req}}^{C^*}$ is shown as black dotted line as the baseline. The line performance of modulation formats is shown as dashed curves (GMI) and solid curves (MI). Focusing on the target at NMI/NGMI $=0.8$, we use circle markers to highlight the position.  As shown by the double-headed arrows, $\Delta\text{SNR}^{R}_{\text{req}}$ is the gap between the black dotted line and the circle markers when NMI or NGMI $=0.8$.
\end{example}

\begin{figure*}[!tb]
    \centering
    \includegraphics[width=\textwidth,height=22em]{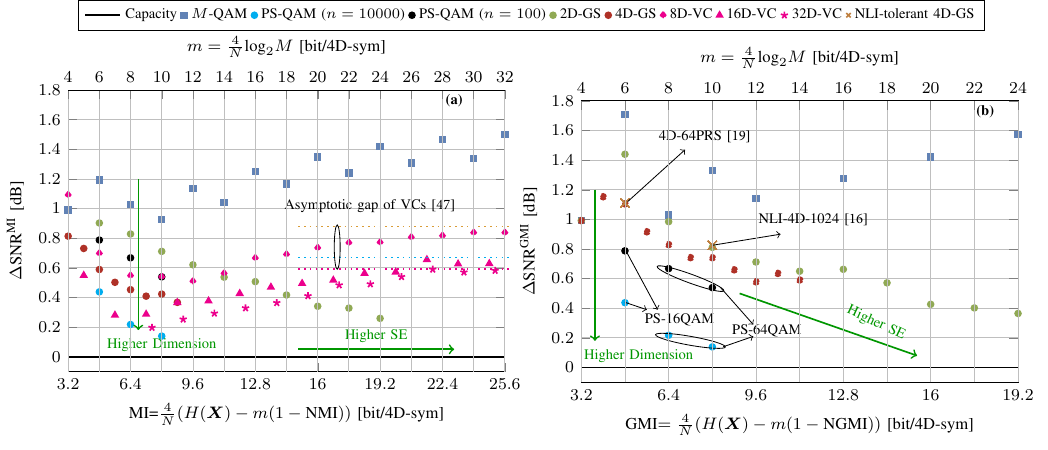}
    \vspace{-1em}
     \caption{\ch{Comparison of the gap to AWGN capacity for different MD  formats at the target rates: (a) NMI=0.8 or (b) NGMI =0.8. Note that the modulation formats includes traditional uniform-QAM, PS-QAM, MD-Vcs and 2D and 4D MI/GMI-optimal modulation formats. Specially, the PS-16QAM with entropy $H(\bX)=$4.8~bit/4D-sym was generated using PM-16QAM ($m=8$~bit/4D-sym), and the PS-64QAM with entropy $H(\bX)=\{8.8, 10.4\}$~bit/4D-sym was generated using PM-64QAM ($m=12$~bit/4D-sym).}}
    \label{fig:CapacityGap_MI}
\end{figure*}

In Fig.~\ref{fig:CapacityGap_MI} (a), the gap to AWGN capacity for various MD formats at the target rate $\text{MI} = 0.8m$ is shown. 
The traditional PM-QAM (\tikz{\draw plot[mark=square*, mark size=2,mark options={color=blue, thick}] (,);)}) shows a zigzag curve.  We can observe that square-QAM constellations tend to have slightly higher SNR gaps, because square-QAM constellations ($m$=4, 8, 12...) generally exhibit a lower shaping gain compared to cross-shaped QAM constellations.  Moreover, as the cardinality of the QAM constellation increases, the SNR gap to AWGN capacity also increases.

For all the MI-optimized 2D-GS (\tikz{\draw plot[mark=*, mark size=2,mark options={color=green, thick}] (,);)}) and 4D-GS constellations (\tikz{\draw plot[mark=*, mark size=2,mark options={color=red, thick}] (,);}) in Fig.~\ref{fig:CapacityGap_MI} (a), it is observed that the SNR gap of the constellations can be brought closer to the Gaussian capacity by increasing the dimension or SE, which has been proved in \cite[Table~\uppercase\expandafter{\romannumeral3}]{ForneyJSAC1984}. In addition, these constellations exhibit a larger shaping gain than the traditional uniform QAM. This can be attributed to the fact that GS changes the constellation positions to mimic a Gaussian distribution. Fig.~\ref{fig:CapacityGap_MI} (a) also shows that the 4D-GS constellations perform better than the 2D-GS constellations. This is because the 4D space can achieve a large minimum Euclidean distance compared to the 2D space.

For MD-VCs (8D-VC \tikz{\draw plot[mark=diamond*, mark size=2.5,mark options={color=magenta, thick}] (,);}, 16D-VC \tikz{\draw plot[mark=triangle*, mark size=2,mark options={color=magenta, thick}] (,);} and 32D-VC \tikz{\draw plot[mark=star, mark size=2.5,mark options={color=magenta, thick}] (,);}), the $\Delta\text{SNR}^{\text{MI}}_{\text{req}}$ decreases as the dimension increases, as shown in Fig.~\ref{fig:CapacityGap_MI} (a). 
As the SE increases, $\Delta\text{SNR}^{\text{MI}}_{\text{req}}$ converges to the asymptotic gap $1.53 - \gamma_{\text{s}}$ [dB] (dotted 
lines), where $\gamma_{\text{s}}$ is the asymptotic shaping gains of the shaping lattices listed in \cite[Table~I]{Li2022}. Compared to 2D-GS and 4D-GS formats, 16D- and 32D-VCs realize lower $\Delta\text{SNR}^{\text{MI}}_{\text{req}}$ between 4 and 14 bit/4D-sym, and provide a good trade-off at high SEs due to their low-complexity encoding and decoding algorithms \cite{Li2022}.

In addition to GS modulation formats, we also consider PS-QAMs with the same SE (MI $=4.8, 6.4, 8$~bit/4D-sym) as baselines, which probabilistically shapes with blocklength $n$ =100 (\tikz{\draw plot[mark=*, mark size=2,mark options={color=black, thick}] (,);)}), 10000 (\tikz{\draw plot[mark=*, mark size=2,mark options={color=cyan, thick}] (,);)}). 
It can be observed that the PS constellations with blocklength $n=10000$  perform better than the 2D-GS and 4D-GS constellations. However, as shown in the PS-QAM ($n=100$) constellations, the superior performance of the PS-QAM ($n=10000$)  constellations vanishes as the blocklength $n$ decreases. For MI between 4~bit/4D-sym and 8~bit/4D-sym, the MD-VCs  outperform PS-QAM constellations with short blocklength ($n$=100), while achieving linear performance similar to the PS-QAM ($n$=10000) constellations. These results show that designing higher-dimensional modulation formats will be likely very competitive in AWGN channel or long-haul nonlinear channels. 
\ch{Similar results have recently been reported in \cite{Sebastiaan2024}. Note that the PS-$M$QAM in \cite{Sebastiaan2024} is implemented by shaping the uniform QAM to achieve a target entropy of $H(\bX) = 0.8 \text{log}_2M$ (i.e., $R^*$=0.8) without considering the redundancy for FEC overhead. 
In this work, the PS-$M$QAM are probabilistically-shaped from higher-order uniform QAM with higher entropy,  and thus the same FEC overhead of 25\% can still be considered in the overall redundancy.} 

{Lastly, we have also presented the gap to AWGN capacity for various formats at the target rate of GMI$=0.8m$, as illustrated in Fig.~\ref{fig:CapacityGap_MI} (b). Note that the 2D-GS and 4D-GS modulation formats are GMI-optimized formats. It also shows that (i) compared to uniform $M$-QAM, MD geometric shaping can provide significant gain, (ii) the SNR gap with a decreasing trend for increasing values of SE or dimension, and (iii) the shaping gains of 4D-GS outperform 2D-GS. In addition, two nonlinearity-tolerant 4D formats (4D-64PRS \cite{BinChenJLT2019} and NLI-4D-1024 \cite{Ling2022}) are highlighted as cross marks \tikz{\draw plot[mark=x, mark size=2.5,mark options={color=brown, thick}](, );}  for GMI of 4.8~bit/4D-sym  and 8~bit/4D-sym in Fig.~\ref{fig:CapacityGap_MI} (b).  These two constellations lead to a small but negligible loss compared to the AWGN-optimized 4D-GS formats. However, more gains are expected in the nonlinear optical fiber channel (see Fig.~\ref{fig:NLI_results} in the next subsection).} For MD-VCs finding efficient labelings that lead to a better performance than QAM at NGMI$=0.8$ is challenging, thus, the $\Delta\text{SNR}^{\text{GMI}}_{\text{req}}$ results for MD-VCs are still an open problem and not shown in Fig.~\ref{fig:CapacityGap_MI} (b).  

\remark{The designing of bandwidth-efficient transceivers can mitigate the band-limited by the fiber loss and operating range of the optical amplifiers \cite{David2010}. Therefore, the designed shaping system needs to work together with the coded modulation (CM) for which the FEC and constellation mapping are designed jointly. Additionally, the AIR, as the cost function for constellation shaping optimization, depends not only on the channel but also on the CM structures (e.g., multilevel codes (MLC) and non-binary FEC, BICM). Therefore, it is necessary to choose different AIRs (MI or GMI) for different systems. Typically, the MI is utilized for the symbol-wise coded modulation system (e.g., MLC-MSD and multi-binary FEC), which is the largest achievable rate for any memoryless channel \cite{AlvaradoJLT2018}. For instance, the authors in \cite{Matsumine2022,Soleimanzade2023JLT} proposed a non-linearity tolerant multi-dimensional GS method based on a two-level MLC  receiver to provide a significant improvement. In \cite{li2023coded}, a MLC-MSD receiver has been used with MD-VC (up to 32 dimensions). On the other hand, the GMI is used for the bit-wise coded modulation system (e.g., BICM and binary FEC), which provides accurate predictions of the performance of CM transceivers when using capacity-approaching SD-FEC decoders. For example, in a study by \cite{ChenJLT2023}, a GMI-based GS modulation optimization was proposed for BICM system with SD-FEC (up to 10~bit/4D-sym).}

\subsection{
Numerical Fiber-Optical Transmission Simulations: Shaping Gains in Nonlinear Optical Fiber Channel}
In this subsection, the optical system under consideration is a 11-channel WDM  system with  96~GBaud symbol rate  and 100~GHz channel spacing. \changed{The considered transmission systems} for evaluating the nonlinear performance of the modulation formats are a 60$\times$80~km multi-span transmission link and a 205~km single-span transmission link \changed{using a standard single-mode fiber.} The fiber parameters are the same as the setting in Sec.~\ref{sec:application_4D_NLI_model}. 
To quickly evaluate the performance for each channel at the given system setting, we estimate the NLI power using the 4D NLI model (Eq.~\eqref{SNR-general2} and Eq.~\eqref{eta}]). 
The selected modulation formats with an SE of 6~bit/4D-sym and 8~bit/4D-sym are listed in Table~\ref{tab:NLI_SNR}. The traditional $M$QAM and PS-$M$QAM with shaping blocklength $n$=10000 are chosen as baseline formats and shown as (\tikz{\draw [solid, color = blue] (1, 1) -- (1.5, 1); \draw plot[mark=square*, mark size=1.5,mark options={color=blue, thick}] (1.25, 1);}, \tikz{\draw [solid, color = cyan] (1, 1) -- (1.5, 1);}). 
The 2D and 4D MI-optimal modulation formats are shown as solid curves (\tikz{\draw [solid, color = green] (1, 1) -- (1.5, 1); \draw plot[mark=*, mark size=1.5,mark options={color=green, thick}] (1.25, 1);}) and (\tikz{\draw [solid, color = red] (1, 1) -- (1.5, 1); \draw plot[mark=*, mark size=1.5,mark options={color=red, thick}] (1.25, 1);}), respectively. Using the same color for the same dimension, the GMI-optimal modulation formats are shown as dashed curves (\tikz{\draw [dashed, color = green] (1, 1) -- (1.5, 1); \draw plot[mark=*, mark size=1.5,mark options={color=green, thick}] (1.25, 1);}, \tikz{\draw [dashed, color = red] (1, 1) -- (1.5, 1); \draw plot[mark=*, mark size=1.5,mark options={color=red, thick}] (1.25, 1);}). Note that GS-AWGN-2D-32 \cite{EricJLT2022} is not only MI-optimal but also GMI-optimal modulation format. 
Furthermore, two nonlinearity-tolerant 4D formats (4D-64PRS \cite{BinChenJLT2019} designed with constraints of constant modulus and NLI-4D-1024 \cite{Ling2022} numerically optimized to GMI for the single-channel single-span scenario at 220~km) are shown as (\tikz{\draw [dashed, color = brown] (1, 1) -- (1.5, 1); \draw plot[mark=x, mark size=1.5,mark options={color=brown, thick}] (1.25, 1);}). 

\begin{table}[!tb]
		\centering
		\caption{Summary of the $\text{SNR}_{\textnormal{req}}^R$ and $\text{SNR}_{\textnormal{eff}}$ of the various constellations in Fig.~\ref{fig:CapacityGap_MI} and Fig.~\ref{fig:NLI_results}.
  }
		\scalebox{0.93}{

\begin{tabular}{c|l|c|c|c}
\hline
\hline
\multirow{2}{*}{SE ($m$)}
 & \hspace{3em} \multirow{2}{*}{Constellation} & $\text{SNR}_{\text{req}}^{\text{MI}}$  & $\text{SNR}_{\text{req}}^{\text{GMI}}$  & $\text{SNR}_\text{eff}$  
\\ &   &  (Fig.~\ref{fig:CapacityGap_MI}) &  (Fig.~\ref{fig:CapacityGap_MI}) & (Fig.~\ref{fig:NLI_results})
\\
\hline
\hline
\multirow{8}{*}{6}
& \textcircled{\tiny{1}} 8QAM (\tikz{\draw [solid, color = blue] (1, 1) -- (1.5, 1); \draw plot[mark=square*, mark size=1.5,mark options={color=blue, thick}] (1.25, 1);})& 7.502 & 8.117 &  10.995\\

\cline{2-5}
& \textcircled{\tiny{2}} hepta2-8 (\tikz{\draw [solid, color = green] (1, 1) -- (1.5, 1); \draw plot[mark=*, mark size=1.5,mark options={color=green, thick}] (1.25, 1);})& 7.215 & 7.830 & 11.090 \\

\cline{2-5}
&\textcircled{\tiny{3}} C4-64 (\tikz{\draw [solid, color = red] (1, 1) -- (1.5, 1); \draw plot[mark=*, mark size=1.5,mark options={color=red, thick}] (1.25, 1);})& 6.901 & 8.835 & 11.078\\

\cline{2-5}
& \textcircled{\tiny{4}} DSQ2-8 (\tikz{\draw [dashed, color = green] (1, 1) -- (1.5, 1); \draw plot[mark=*, mark size=1.5,mark options={color=green, thick}] (1.25, 1);}) & 7.332 & 7.752 & 11.005 \\

\cline{2-5}
&\textcircled{\tiny{5}} GS-AWGN-4D-64 (\tikz{\draw [dashed, color = red] (1, 1) -- (1.5, 1); \draw plot[mark=*, mark size=1.5,mark options={color=red, thick}] (1.25, 1);}) & 7.242 & 7.417 & 10.962 \\


\cline{2-5}
&\textcircled{\tiny{6}} 4D-64PRS (\tikz{\draw [dashed, color = brown] (1, 1) -- (1.5, 1); \draw plot[mark=x, mark size=2,mark options={color=brown, thick}] (1.25, 1);}) & 7.257 & 7.421 & 11.166 \\

\cline{2-5}
& \textcircled{\tiny{7}} PS-16QAM($n=10^4$)(\tikz{\draw [color = cyan] (1, 1.2) -- (1.5, 1.2);}) & 
\multicolumn{2}{c|}{${\text{SNR}_{\text{req}}^{R}}=6.75$}  & 10.741\\

\cline{2-5}
& Gaussian modulation & 
\multicolumn{2}{c|}{$\text{SNR}_{\text{req}}^{C}=6.312$}  & 10.705\\

\hline
\multirow{6}{*}{10}
&\textcircled{\tiny{8}} 32QAM (\tikz{\draw [solid, color = blue] (1, 1) -- (1.5, 1); \draw plot[mark=square*, mark size=1.5,mark options={color=blue, thick}] (1.25, 1);})& 12.686 & 13.091 & 13.119\\


\cline{2-5}
&\textcircled{\tiny{9}} GS-AWGN-2D-32 (\tikz{\draw [dashed, color = green] (1, 1) -- (1.5, 1); \draw plot[mark=*, mark size=1.5,mark options={color=green, thick}] (1.25, 1);}) & 12.472 & 12.572 & 12.841\\


\cline{2-5}
&\textcircled{\tiny{10}} C4-1024 (\tikz{\draw [solid, color = red] (1, 1) -- (1.5, 1); \draw plot[mark=*, mark size=1.5,mark options={color=red, thick}] (1.25, 1);})& 12.184 & 14.614 & 12.846\\

\cline{2-5}
&\textcircled{\tiny{11}} 4D-OS1024 (\tikz{\draw [dashed, color = red] (1, 1) -- (1.5, 1); \draw plot[mark=*, mark size=1.5,mark options={color=red, thick}] (1.25, 1);}) & 12.291 & 12.502 & 12.767\\

\cline{2-5}
&\textcircled{\tiny{12}} NL-4D-1024 (\tikz{\draw [dashed, color = brown] (1, 1) -- (1.5, 1); \draw plot[mark=x, mark size=2,mark options={color=brown, thick}] (1.25, 1);})& 12.417 & 12.586 & 13.085\\

\cline{2-5}
& \textcircled{\tiny{13}} PS-64QAM($n=10^4$)(\tikz{\draw [color = cyan] (1, 1.2) -- (1.5, 1.2);}) & 
\multicolumn{2}{c|}{${\text{SNR}_{\text{req}}^{R}}=11.9$}  & 12.265\\

\cline{2-5}
& Gaussian modulation &\multicolumn{2}{c|}{$\text{SNR}_{\text{req}}^{C}=11.761$}  
 & 12.143 \\
\hline
\hline
\end{tabular}
		}
  \label{tab:NLI_SNR}
\end{table}

\begin{figure}[!tb]
    \centering
    \includegraphics[width=.5\textwidth,height=17.5em]{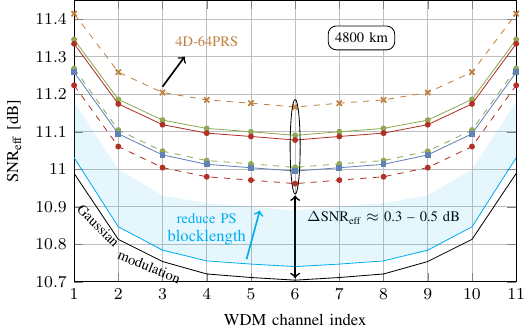}
    \includegraphics[width=.5\textwidth,height=17.5em]{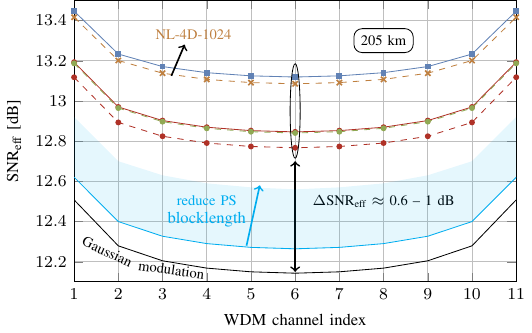}
    \caption{The effective SNR performance comparison  over two different transmission cases via 4D NLI model. The considered constellations are shown in Table~\ref{tab:NLI_SNR}.
    }
    \label{fig:NLI_results}
  \vspace{-1em}
\end{figure}

\begin{figure}[!tbp]
    \centering
    \includegraphics[width=.5\textwidth,height=16em]{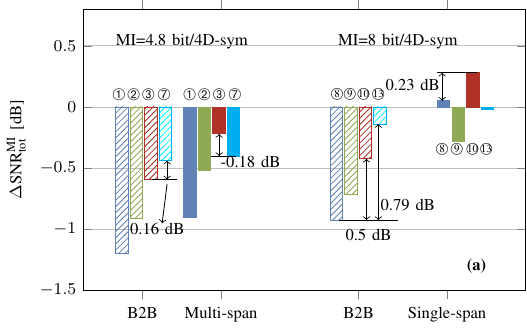}
    \includegraphics[width=.5\textwidth,height=16em]{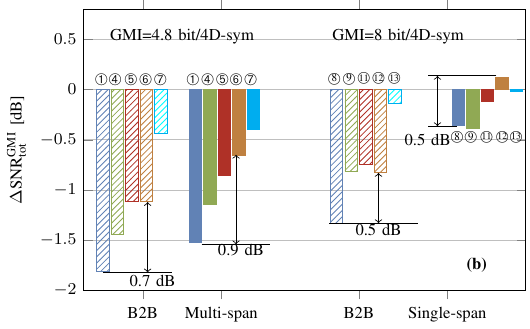}
    \caption{The relative shaping gap or gain  in terms of SNR 
    for  the considered constellations in Table~\ref{tab:NLI_SNR} (marks as \textcircled{\scriptsize{1}} -- \textcircled{\scalebox{0.9}{\scriptsize{13}}}) with respect to Gaussian   distribution.
    }
    \label{fig:relative_gains}
    \vspace{-1em}
\end{figure}

Table~\ref{tab:NLI_SNR} shows the required and effective SNR differences for the modulation formats with an SE of 6~bit/4D-sym and 10~bit/4D-sym, respectively. \changed{ 
To target on a practical SD-FEC around 25\% overhead, we show the required SNRs (i.e., $\text{SNR}_{\text{req}}^{\text{MI}}$ and $\text{SNR}_{\text{req}}^{\text{GMI}}$) of $R=0.8m$  for AWGN channel.} To assess the effective SNR for two different transmission cases, the modulation formats with SE of 6~bit/4D-sym and 10~bit/4D-sym were simulated over multi-span system and single-span system, respectively. \changed{The $\text{SNR}_{\text{req}}^C$ and $\text{SNR}_{\text{eff}}$  values of Gaussian modulation is obtained via Shannon formula and GN model, respectively.} In the following, we show the optimal effective SNR of each channel and  the relative gains for these modulation formats in Fig.~\ref{fig:NLI_results} and Fig.~\ref{fig:relative_gains}, respectively.

Fig.~\ref{fig:NLI_results} (a) shows the optimal effective SNR of each channel for modulation formats with an SE of 6 bit/4D-sym over $60\times80$~km transmission system. The simulation results in Fig.~\ref{fig:NLI_results} (a) show that the geometrically shaped 4D constellation for the nonlinear fiber channel (C4-64) obviously outperforms PM-8QAM. Since the design of GS-AWGN-4D-64 does not consider modulation-dependent NLI, its effective SNR is slightly worse than PM-8QAM, although it has a great linear shaping gain. Therefore, despite the 2D-GS and 4D-GS designed for the AWGN channel having remarkable required SNR gains, their effective SNR is significantly reduced due to the impact of nonlinear impairments caused during propagation. In addition, we also show the constellation design for a nonlinear fiber channel with constant modulus, which has fewer energy variations. It can be observed that the constant-modulus constellation provides better nonlinearity tolerance than all other constellations (up to 0.5~dB gain compared to Gaussian modulation).

In addition to MD modulation formats, PS-16QAM ($n=10000$) is also considered as a basline in Fig.~\ref{fig:NLI_results} (a). The result shows that the PS-16QAM with shaping blocklength $n=10000$ has the worst effective SNR than other modulation formats. It indicated that PS with blocklength $n=10000$ will have higher penalty in nonlinear fiber channel, and lead to a reduced effective SNR. The results of PS with short-blocklength are not shown here because the NLI model is only suitable for independent symbols. In the future, we will follow the work in \cite{9464637} to extend the 4D NLI model with energy dispersion index (EDI). \changed{However, from the work in \cite{8815813,9083652,9849911,9716753}, we know that short blocklength DM can effectively reduce NLI. Therefore, we can predict that the performance of PS-16QAM increases as the blocklength $n$ decreases for the optical fiber channel. Note that as the block length decreases, the rate loss increases. To further enhance the nonlinear performance of PS schemes, several nonlinearity optimized PS schemes were proposed, such as short block length sphere shaping and sequence selection based sphere shaping, which make a trade-off between nonlinear gain and rate loss.}

In Fig.~\ref{fig:NLI_results} (b), the modulation formats with 8 bit/4D-sym over a 205~km single-span system are shown. As expected, NL-4D-1024 which is numerically optimized in the optical fiber channel  shows 0.94~dB gain compared to Gaussian modulation.  The other modulation formats shows the effective SNR difference between 0.6-0.7~dB, in particular the PS-QAM ($n=10000$) is the worst. It also indicated that the nonlinear interference inhibits shaping gains, especially for the constellations optimized in the AWGN channel. However, the nonlinearity-tolerant geometrically-shaped formats (e.g., NL-4D-1024) exhibit excellent nonlinear tolerance. \changed{In other words, optimized modulation formats for nonlinear optical fiber channel should strike a balance between linear shaping gain and nonlinear SNR loss to achieve the maximum shaping gain,  instead of  only  improving SNR.}

\begin{figure*}[!tb]
    \centering
    \includegraphics[width=\textwidth]{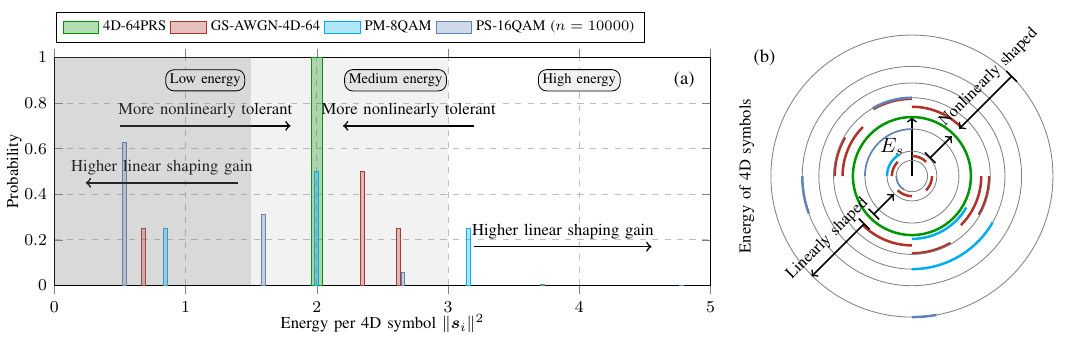}
 \vspace{-1.5em}
    \caption{(a) Probability of energy per 4D symbol for four SE$ = 6$ bits/4D-sym modulations; (b) 2D geometric representation of energy per 4D symbol.}
    \label{fig:Energy}
    \vspace{-1em}
\end{figure*}

To conclude this section and further quantify the gains offered by 2D and 4D shaped constellations in nonlinear optical fiber channel, we propose a relative shaping gains in terms of the SNR ($\Delta \text{SNR}^{\text{MI}}_{\text{tot}}$ and $\Delta \text{SNR}^{\text{GMI}}_{\text{tot}}$) for the considered 2D and 4D modulation formats with respect to Gaussian modulation for both multi-span and single-span scenarios are shown in Fig.~\ref{fig:relative_gains}. 
The total SNR gap or gains considering both linear and nonlinear effects  are defined as $\Delta\text{SNR}^R_{\text{tot}}=-\Delta \text{SNR}^R_{\text{req}}+\Delta \text{SNR}_{\text{eff}}$, where  $\Delta \text{SNR}_{\text{eff}}$    represent the modulation-dependent effective SNR ($\text{SNR}_{\text{eff}}$) difference between a given constellation and Gaussian constellation itself in Table~\ref{tab:NLI_SNR}, respectively.

In Fig.~\ref{fig:relative_gains}, the relative shaping gains are shown for modulation formats with a SE of 6 bit/4D-sym and 10 bit/4D-sym over B2B and optical fiber systems. Fig.~\ref{fig:relative_gains} (a) and (b) show the $\Delta \text{SNR}^{\text{MI}}_{\text{tot}}$ for MI-optimal modulation formats and $\Delta \text{SNR}^{\text{GMI}}_{\text{tot}}$ for the GMI-optimal modulation formats, respectively. We identified these formats via serial number (see Table.~\ref{tab:NLI_SNR}). The traditional QAM formats are chosen as the baselines. \changed{We summarize the results and observations in terms of MI and GMI in Fig.~~\ref{fig:relative_gains} (a) and (b) as follows.}
\begin{itemize}
    \item It can be  obviously observed that the 4D constellations provide positive gains over uniform QAM, but 2D constellation even has negative penalty in GS-AWGN-2D-32 (\textcircled{\footnotesize{9}}). \changed{It indicates that the 4D constellations outperform 2D constellations for both MI and GMI due to the larger dimension space to achieve shaping gains.}
    \item \changed{The results show that the PS constellations has exhibits significant linear shaping gain for both MI and GMI in the B2B case. For example the PS-16QAM with blocklegth $n=10000$ (\textcircled{\footnotesize{7}}) exhibits  a positive gain (0.16~dB) compared to C4-64 (\textcircled{\footnotesize{3}}) in a B2B case due to the Gaussian-shaped distribution.} 
    \item \changed{But the PS constellations are not the optimal choice in optical fiber channel with high nonlinearity. Because  a Gaussian-shaped modulation formats will lead to a large SNR penalty, especially  in optical fiber channel with strong nonlinearity.
    For example, the PS-16QAM with blocklegth $n=10000$ (\textcircled{\footnotesize{7}})  shows negative gain (-0.18~dB) in a multi-span case  compared to C4-64 (\textcircled{\footnotesize{3}}).  In Fig.~7 (b), although the PS-16QAM ($n=10000$) outperforms 4D-64PRS in multi-span case, this conclusion still holds. Because the gap between PS-16QAM ($n=10000$) and 4D-64PRS is reduced by about 0.4~dB and in the case of stronger nonlinearity, we predict that 4D-64PRS can get more gain.}
    \item The nonlinearity-tolerant 4D constellations can maintain or even provide larger SNR gains in optical fiber channel than the gains in B2B cases. For example the 4D-64PRS (\textcircled{\footnotesize{6}}) provides 0.7~dB SNR gain compared to 8QAM (\textcircled{\footnotesize{1}}) in B2B case, but the SNR gain up to 0.9~dB in multi-span cases. The NL-4D-1024 (\textcircled{\footnotesize{12}}) maintains the same SNR gains (0.5~dB) compared to 32QAM. (\textcircled{\footnotesize{8}}) across both scenarios. 
    \item For all constellations shown, they can reduce the gap to the theoretical AWGN-optimal Gaussian modulation in optical channel due to the largest NLI and lowest effective SNR of Gaussian modulation.
    \item Remarkably, no existing modulation format  in the literature is  universally optimal for different types of channels.  For high nonlinear channel, the MD shaping is an efficient way to search for constellations which could balance the tolerance for the linear noise and nonlinear noise. 
\end{itemize}

In order to clearly show the relation between modulation-dependent nonlinearity-tolerance and the symbol's energy, we use a similar analysis approach in \cite{BinECOC2020}. As shown in  Fig.~\ref{fig:Energy}, an example of comparing the  probability distribution of 4D symbol's energy for four SE$ = 6$~bit/4D-sym modulation formats was shown. Here  four constellations were considered (normalized to unit energy per polarization): 4D-64PRS, GS-AWGN-4D-64, PM-8QAM and PS-16QAM with blocklength $n=10000$. As shown in Fig.~\ref{fig:Energy}, the 4D constellation symbols' energy were divided as three groups: low energy, medium energy and high energy. \changed{We can observe that nonlinear shaping is generally in contradiction with linear shaping, which moves the constellation symbols away from the average energy $E_s$. Specially, since the GMI-based shaping optimization is highly non-convex, the linear performance between GS-AWGN-4D-64 and 4D-64PRS is similar. However, nonlinear constraints are introduced to optimize performance while preserving linearity (e.g., constant-modulus for 4D-64PRS). Additionally, 8-QAM which shows a more separated energy distribution has worse linear performance than 4D-64PRS. This is primarily due to its significantly higher number of pairs at minimum squared Euclidean distance compared to Gray-labeled 4D-64PRS and non-Gray-labeled 8QAM \cite{BinChenJLT2019}. 
Therefore, in specific channel scenarios, it is possible to balance the nonlinear and linear shaping gains to maximize the overall gain.}

\section{Physical Realizations and Digital Signal Processing for MD modulation formats}\label{sec:discussion}
It is important to note that the DOFs used as signaling dimensions were typically assumed to be independent noise sources with each other. However, due to the nonlinear and dispersion effects in actual optical fibers, signals interact during the transmission process, leading that the noise crosstalk is correlated. These influences make the DOFs no longer independent, making the design and optimization of multidimensional modulation more complex. In this subsection, we briefly discuss the practical implementation challenges by increasing the dimensionality via the physical DOFs. 
\begin{itemize}
    \item Quadratures: One problem is the ``limited scalability".  \changed{Unlike time- and space-related dimensions, quadratures can only  be decomposed into two degrees of freedom, i.e., amplitude and phase, which limits the potential for expanding the capacity of optical communication systems.} Additionally, the optical phase requires a coherent or differential-phase receiver for reliable detection. Despite for now the coherent receivers are becoming more prevalent and cost-effective, simplifying coherent receiver designs is crucial for widespread adoption and efficient utilization of optical phase modulation techniques.
    \item Polarizations: Note that the combined use of quadratures and polarization leads to a  real 4D constellation space for coherent signaling. \changed{However, there are still existing many challenges}, for example the slowly drifting of  polarization state and polarization mode dispersion (PMD). Therefore, the use of polarization dimensions requires the polarization tracker and adaptive constant modulus algorithm filter to resolve the polarization state drifting  and PMD, respectively.
    \item Time slots: In order to ensure that the receiver can  correctly parse and synchronize the received data, higher precision frame and clock synchronization are required. And the introduction of channel estimation and equalization techniques to address the variations in signal transmission characteristics across different time slots, this increases the complexity of DSP at receiver. On the other hand, the temporal dimension provides a simple playground for testing new formats without challenging synchronization issues, as demonstrated by van Uden et al.  in \cite{Uden2014OFC} using two time slots to form a 4D symbol instead of two polarization states, significantly simplifying receiver DSP.
    \item Wavelengths: Multidimensional supersymbols can be formed using different channels transmitting at adjacent wavelengths, known as ``superchannels".  However, utilizing correlated modulation to increase signal space dimensionality with superchannels has not yet been achieved. Joint detection of multi-wavelength channels in a WDM link could enable multidimensional modulation and coding, but practical problems with temporal and phase synchronization and signal ambiguities have limited its use. Nonetheless, a few limited cases have been reported, such as the 8D format by Eriksson et al. \cite{ErikssonECOC2013}, which detected two wavelengths using the same local oscillator.
    \item Space (Cores/modes): Spatial-division multiplexing has been used to increase data rates of optical links, and further increasing capacity by using joint transmission over parallel modes and/or fibers.  However, there are many practical challenges such as  differential mode delay and modal crosstalk. In \cite{Uden2014OFC}, Relevant scholars used MIMO (multiple input multiple output) signal processing to  compensate for the modal crosstalk, but it is very DSP-heavy. Using different (single-mode) cores in a multicore fiber seems to be a more straightforward approach, which relaxes the crosstalk and walk-off penalties. Novel formats exploiting these DOFs have been discussed by Eriksson et al. \cite{Eriksson2014JLT} and some initial experiments were reported by Puttnam et al. \cite{Puttnam2014OE, Puttnam2014}. 
    \changed{Recently, a MD Voronoi constellation with concatenated two-level MLC coded modulation scheme were proposed for multi-core fiber communication system to  achieve better performance–complexity trade-off \cite{Zhao2024}, and also demonstrated by a four-core experimental transmission \cite{Zhao2024ECOC}.}
\end{itemize}

\changed{We now discuss the challenges associated with using multidimensional modulation versus  PS  schemes. 
Conventional PS schemes are typically based on the processing of one-dimensional signals.  As a result, they generally require only a single mapper (or interleaver) to achieve a wide range of transmission rates (or to mitigate the impact of burst errors on signal transmission).  In contrast, multi-dimensional modulation formats must handle signals across multiple dimensions, necessitating multiple distinct mappers (or interleavers) to accommodate such diverse rates (or to manage higher-dimensional data arrangements). Furthermore, due to the irregular constellation point distribution of MD-GS, more precise digital-to-analog converters (DACs) and analog-to-digital converters (ADCs) are required \cite{ PfauJLT2009}, significantly increasing complexity.}

\changed{On the other hand, while conventional PS schemes only require an external distribution matcher with lower complexity, it limits the shaping gain. Specially, For the temporal shaping PS with short blocklength,  twice the amount of interleavers must be used with PS to avoid burst errors induced by the temporal structure of the PAS block and the nonlinear interference \cite{9083652}. In contrast, the MD-GS scheme only needs simple modifications to the mapper and demapper, enabling easy integration with FEC and customization for various impairments \cite{8535199,9224128}. }

\section{Conclusion}\label{sec:conclusion}
The contributions of this manuscript are threefold. First, the main contribution of this manuscript is an analytical NLI model-based shaping gain estimation method is proposed. In particular, our approach is based on the 4D NLI model to rapidly estimate the performance of arbitrary MD modulation formats. Secondly, we further extended the analytical 4D NLI model proposed in our previous work \cite{Liang2023JLT} to apply to PS constellations and derived a simple analytical expression to analyze the characteristics of the NLI. Thirdly, we offer a comprehensive analysis of various MD modulation formats (including GS and PS) using this method and present relevant analysis and conclusions.

The results suggest that more gains could be harvested with high-dimensional space for the AWGN channel and more nonlinear tolerance could be found with four-dimensional space for the nonlinear optical channel. Additionally, MD-GS constellations outperform PS in the optical fiber channel with strong nonlinearity.
Therefore, high-dimensional space design
could be the key to unlocking the full potential of nonlinear constellation shaping.

\bibliographystyle{IEEEtran}
\bibliography{references}


\end{document}